# Synthesis and spectroscopic characterization of completely isotactic polyacrylonitrile

**Jun-Ting Zou, Yu-Song Wang, Wen-Min Pang, Lei Shi,\* and Fei Lu**

Hefei National Laboratory for Physical Sciences at Microscale, University of Science and Technology of China, Hefei, Anhui 230026, China

\* Corresponding author. Tel.: +86-551-3607924; E-mail addresses: shil@ustc.edu.cn

**Abstract:**

Completely isotactic polyacrylonitrile (i-PAN) has been synthesized successfully by an improved urea inclusion polymerization. The tacticity of prepared samples were confirmed by $^{13}$C nuclear magnetic resonance (NMR), Fourier transform infrared (FT-IR) and X-ray diffraction (XRD).

## Introduction

Polyacrylonitrile (PAN) is one of the most widely used polymers for textiles and precursors of carbon fibers.[1] The latter are particularly suitable to produce high performance polymer matrix composites characterized by high strength, stiffness, and lightweight.[2] Commercially available PAN is usually synthesized by radical polymerization without specific stereoregularity along the chain direction.[3] Stereoregular polymers often exhibit particular and valuable properties, which has been attracting much attention both in academic and industrial fields. It is realized that the isotacticity of PAN is likely one of the important factors affecting the performance of carbon fibers.[4, 5] Thus, a complete isotactic acrylonitrile (AN) polymer is a good candidate for making high-quality carbon fibers.[6-8] A number of methods have been tried in order to produce PAN in higher isotactic level,[9, 10] but only the radiation-induced inclusion polymerization in urea (urea canal polymerization) is proven to be the most effective way.[11] In 1992, Minagawa et al.[11] made a comprehensive investigation of the experimental conditions leading to PAN prepared in the urea canal complex, which focused on producing more highly isotactic PAN. They were able to prepare PAN samples, with preponderantly isotactic more than 80% mm-triads. However, up to now, completely isotactic PAN (mm>99%) has not yet been obtained.

In present work, completely isotactic PAN (i-PAN) was prepared by radiation-induced inclusion polymerization successfully. $^{13}$C nuclear magnetic resonance (NMR), X-ray diffraction (XRD), and Fourier transform infrared (FT-IR) were used to characterize the prepared samples.

## Experimental

*Materials.* Acrylonitrile was vacuum distilled followed by fractional distillation over $CaH_2$ under nitrogen and reserved with molecular sieve (4-A). Urea was purified by recrystallization, and other chemicals were used without further purification.

*Sample Preparation.* The process of the completely i-PAN synthesis can be divided into four elementary steps: (1) acrylonitrile/urea inclusion compounds (AN/UIC) formation, (2) γ-ray irradiation, (3) chain propagation (post-polymerization), and (4) recovery. The AN/UIC were prepared by mixing AN with urea at the molar ratio 1 : 2 (AN : urea) under vacuum, followed by cooling to a low temperature (-50 °C) and keeping for a long time (200 h) to ensure that AN monomers were included in the canals of urea totally. After aging, the prepared AN/UIC were cooled down to liquid nitrogen temperature (-196 °C), and irradiated under a $2.0 \times 10^{15}$ Bq $^{60}$Co γ-ray source at a dose rate about 80 Gy min$^{-1}$ for 4 h. In the process of the chain propagation, the post-polymerization proceeded at -100 °C without γ-ray irradiation, and the nitrogen with the same temperature was inputted to remove the heat of polymerization. After post-polymerization for 5 h, the polymer-urea mixture was crushed in cooled methanol. The residue was washed repeatedly with distilled water and methanol until confirmation of complete removal of the unreacted monomers and urea. Finally, the polymer was dried in vacuum, obtaining i-PAN. For comparison, PAN with a content of 77% isotactic triads (h-PAN) was prepared in the manner by the method proposed by Minagawa *et al*.[11] and the atactic PAN (a-PAN) was synthesized by solution polymerization.[12] The polymerization results and characteristics of the obtained samples are shown in Table 1.

*Measurements*. The $^{13}$C NMR experiments reported in this study were carried out on a Buker Avance 400 NMR spectrometer at 60℃ under the following operating conditions: inverse gated decoupling mode; relaxation delay time 6s; acquisition time 0.541s; pulse time 13.40μs; accumulation, more than 4000 times. PAN samples were dissolved in DMSO-$d_6$ with concentration of 10 wt% and tetramethylsilane was used as internal standard (0 ppm). The gel permeation chromatography determinations were performed by using a GPC system consisting of a Waters 1515 HPLC pump, four Waters Styragel columns (HR1, HR3, HR4, HR5), and a Waters 2414 refractive index detector. The columns system were thermostated at 60 °C. The eluted carrier solvent was DMF with NaNO$_3$ (0.065 mol/L) to eliminate the electrostatic effect and flow rate was 1 ml/min. The calibration to determine the molecular weight of the synthesized polymers was performed, using a polyacrylonitrile standard (from Aldrich) with $M_w$ of 86,200 and $M_n$ of 22,600. FTIR spectra were obtained using KBr pellets on a Nicolet 8700 FTIR spectrometer. X-ray diffraction (XRD) patterns were measured on a Rigaku TTR III diffractometer with Cu Kα radiation source (λ = 1.54187 Å).

## Results and discussion

It is believed that an ideal isotactic configuration can be attained only within the urea canals, and AN monomers outside the tunnels can be polymerized as atactic PAN. Hence, the prerequisite of the completely isotactic PAN synthesis is that AN monomers should be included in the canals of urea totally at the AN/UIC formation step. Since considerable heat is released in polymerization, and the thermal conductivities of the urea and PAN are poor, the local temperature elevation produced by the heat of polymerization will destroy inclusion compounds, decreasing the isotacitcity of polymer products. Thus, the removal of heat of polymerization cannot be ignored so as to obtain i-PAN.

High resolution $^{13}$C NMR has been confirmed to be an effective tool for determining the stereoregularity of polymer chains.[13] The typical $^{13}$C NMR spectra of i-PAN and h-PAN are shown in Fig. 1. The differences in stereoregularity appear in the nitrile (CN) carbon, methylene ($CH_2$) carbon and methine (CH) carbon in the NMR spectra. Single peak appears in the region of CN carbon, $CH_2$ carbon and CH carbon in the $^{13}$C NMR spectrum of i-PAN, which indicates that there is only one stereo configuration along the chain direction within the limit of experimental detection. J. Schaefer firstly reported $^{13}$C NMR spectra of PAN and assigned the nitrile carbon spectra with triads.[14] According to his results, the assignment of the triad tacticity is made in Fig. 1. It can be reasonably inferred that the single peak in the region of CN carbon of i-PAN should be assigned to the isotactic sequence. On the other hand, according to the assignments made by Wang *et al.*[15], in the nitrile carbon spectra of i-PAN, h-PAN and a-PAN shown in Fig. 2, the peak at 120.41 ppm should be assigned to the (mmmmmmmm) or even longer isotactic sequence of PAN. In other words, the peak at 120.41 ppm represents long isotactic sequences no shorter than the ennead. For i-PAN, there is only the isotactic configuration along the whole chain. Prudently considering the detection limit of $^{13}$C NMR, the triad isotacticity (mm) is more than 99%.

The FTIR spectra of three samples are shown in Fig. 3, which are similar to the results reported by M. Minagawa et al.[16] The differences in stereoregularity appear in the IR bands in the region of 1300-1200 and 540 cm$^{-1}$. Minagawa et al.[16, 17] suggested that there is a quantitative relationship between the relative intensity of bands at 1300-1200 cm$^{-1}$ (D1230/D1250) and the content of isotactic triad units. However, in present study, the baselines in the region of 1300-1200 cm$^{-1}$ are so ambiguous that the quantitative relationship cannot be applied. From a point of non-quantitative view, when the isotacticity is increased, the intensity of the 1230-cm$^{-1}$ band is enhanced.

Other IR spectroscopic features of stereoregular PAN appear in the absorption band near 540 cm$^{-1}$, that is, when the isotacticity of PAN is increased, the peak intensity is enhanced and the full width at half maximum (FWHM) decreases (Fig. 3). As shown in Fig. 4, it is apparent that the FWHM of the 540-cm$^{-1}$ peak is directly related to the content of isotactic triad units in PAN. For i-PAN, the FWHM of the 540-cm$^{-1}$ peak is about 17 cm$^{-1}$, which is coincident with the prediction by Minagawa et al.[16]

Fig. 5 shows the XRD patterns of i-PAN, h-PAN and a-PAN. It is found that when the extent of the stereoregularity was increased, the main peak at about 17 degree (2θ) was sharpened and shifted from 17.1 to 16.77°. Minagawa et al.[18] carried out a detailed wide-angle X-ray diffraction study for a series of stereoregular PAN with different isotacticities (83-25%), and predicted that the values corresponding to a 100% isotactic sample are: 5.27$_5$Å (intermolecular distance) and 0.97° (half-value width of the main peak). As shown in Fig. 5, the intermolecular distance is 5.2822 Å, and the half-value width of the main peak is 0.96°, which reveals that the PAN sample is a completely isotactic PAN (mm>99%).

## Conclusions

In summary, we have successfully synthesized the completely isotactic PAN

(mm>99%) by radiation-induced inclusion polymerization. $^{13}$C nuclear magnetic resonance (NMR), X-ray diffraction (XRD), and Fourier transform infrared (FT-IR) were used to characterize the prepared samples. More detailed properties of i-PAN are under investigating.

Table 1 The polymerization results and characteristics of the samples prepared in this study.

| Codes | Polymerization | $M_n^a$ | $M_w^a$ | Polydispersity[a] | tacticity | | |
|---|---|---|---|---|---|---|---|
| | | | | | mm | mr | rr |
| i-PAN | present work | 45469 | 59884 | 1.32 | >0.99 | ~0 | ~0 |
| h-PAN | ref. 11 | 62546 | 92380 | 1.48 | 0.77 | 0.18 | 0.05 |
| a-PAN | radical | 74867 | 126674 | 1.69 | 0.26 | 0.50 | 0.24 |

[a] Obtained from GPC. $M_n$: number average molecular weight; $M_w$: weight average molecular weight

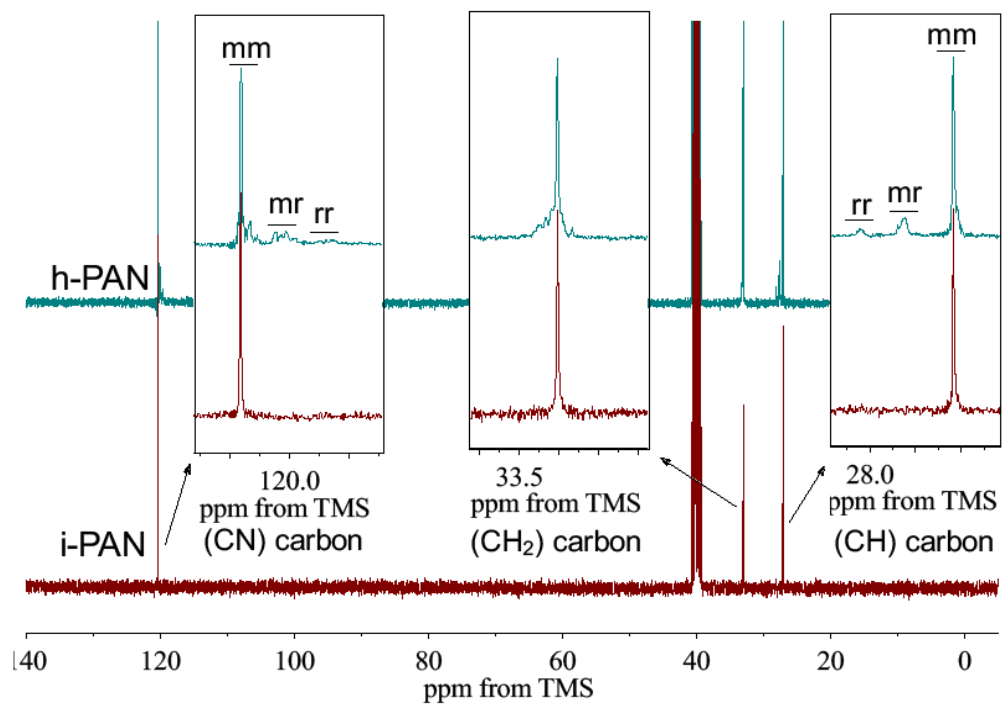

Fig. 1. The typical $^{13}$C NMR spectra of h-PAN and i-PAN.

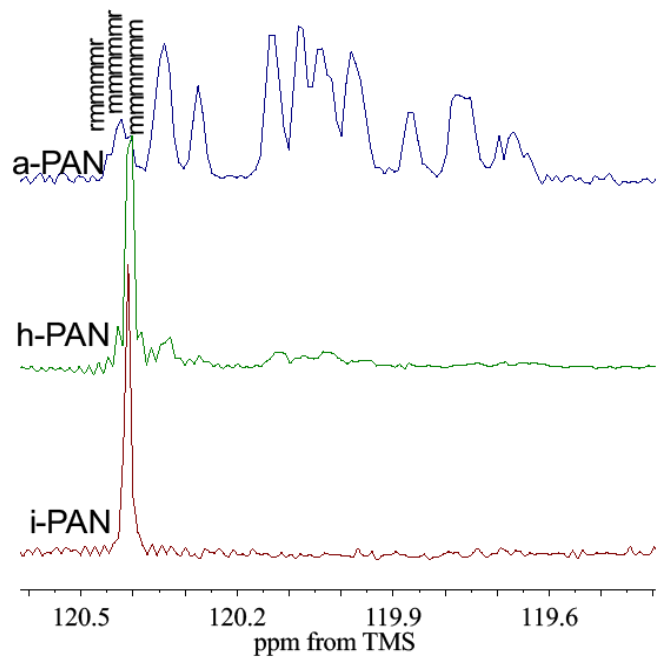

Fig. 2. $^{13}$C NMR CN peak spectra of i-PAN, h-PAN and a-PAN.

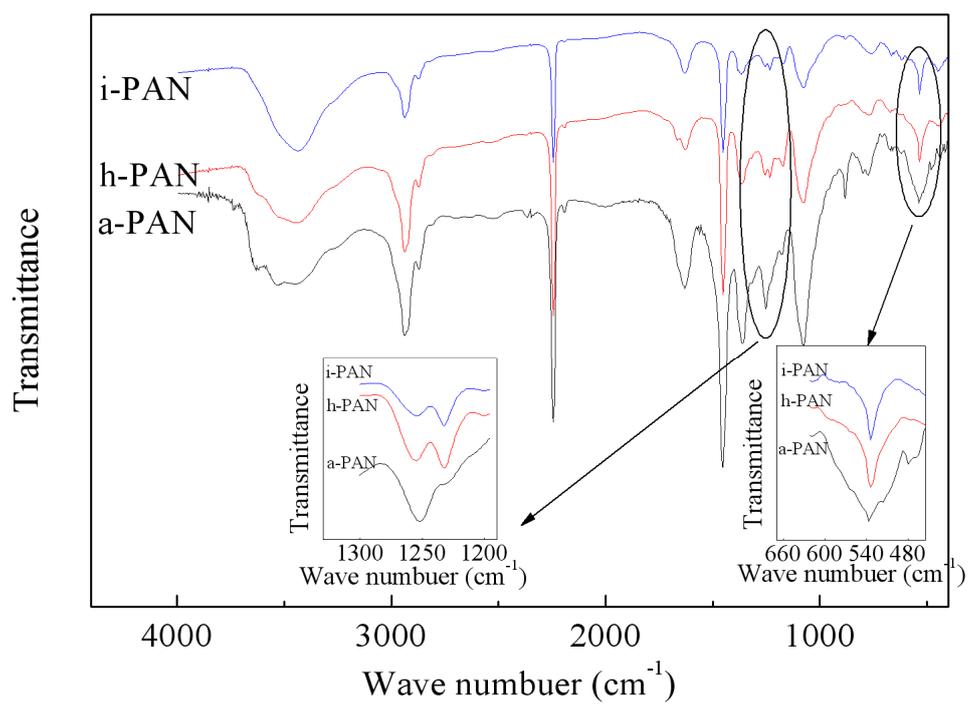

Fig. 3. FTIR spectra of i-PAN, h-PAN and a-PAN.

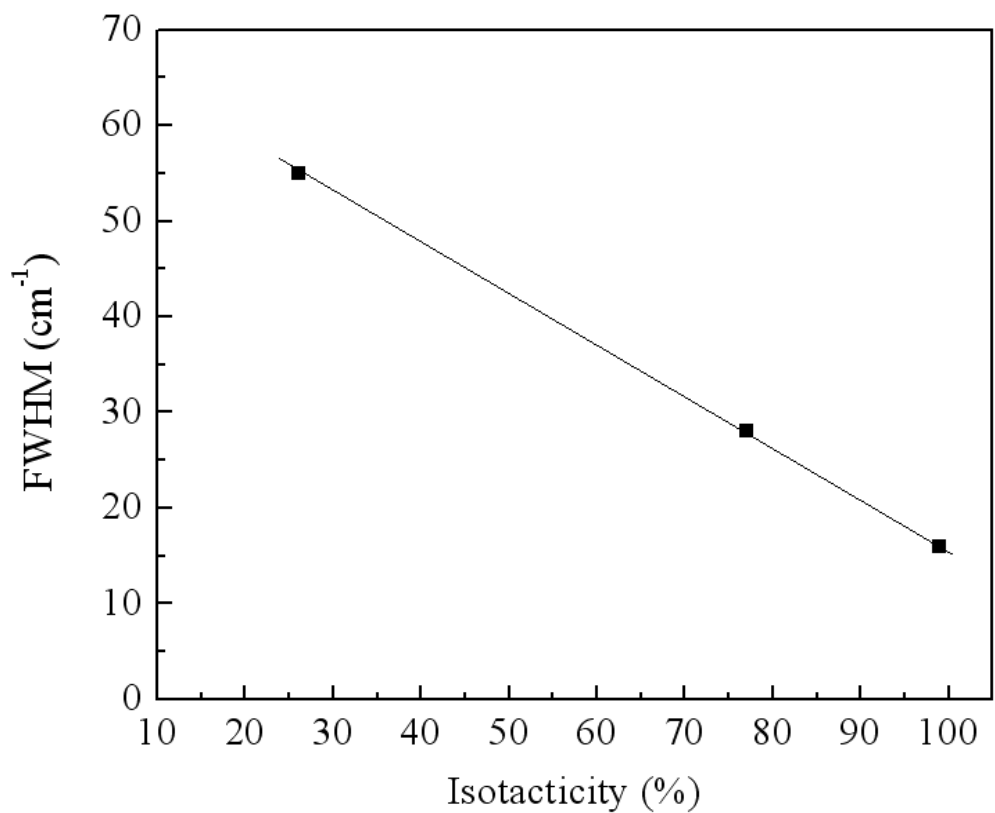

Fig. 4. FWHM of the 540-cm$^{-1}$ peak versus stereoregularity of PAN.

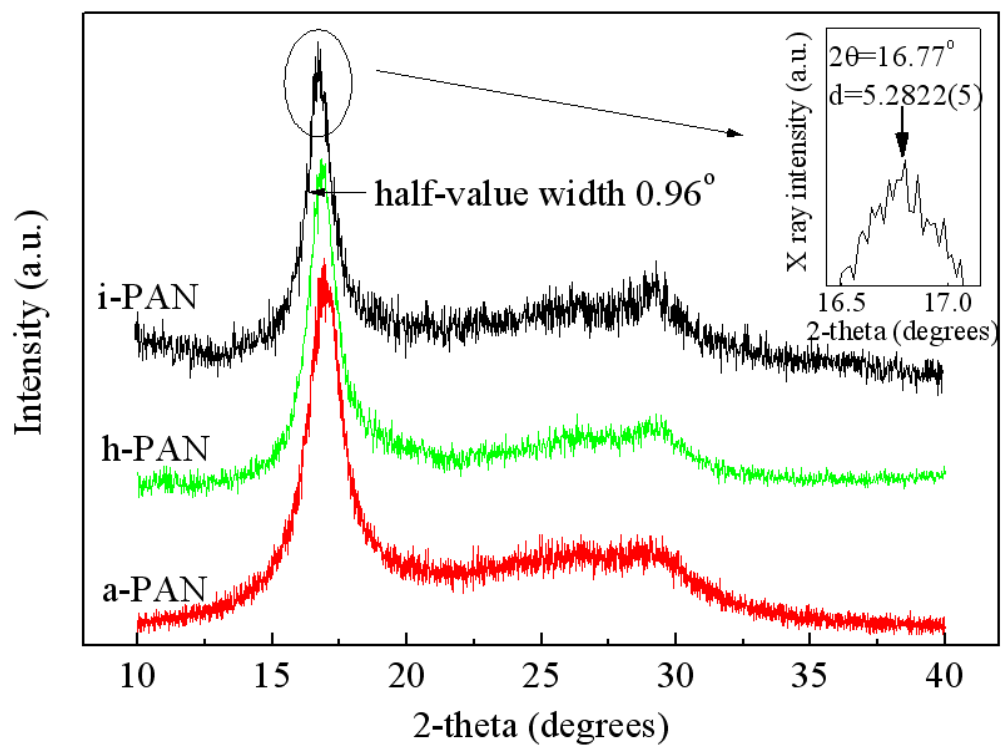

Fig. 5 XRD patterns of i-PAN, h-PAN and a-PAN